# An Electromagnetic-Information-Theory Based Model for Efficient Characterization of MIMO Systems in Complex Space

Ruifeng Li, Da Li, *Member, IEEE*, Jinyan Ma, Zhaoyang Feng, Ling Zhang, *Member, IEEE*, Shurun Tan, *Member, IEEE*, Wei E. I. Sha, Senior *Member, IEEE*, Hongsheng Chen, *Fellow, IEEE,* and Er-Ping Li, *Fellow, IEEE*

*Abstract*—It is the pursuit of a multiple-input-multiple-output (MIMO) system to approach and even break the limit of channel capacity. However, it is always a big challenge to efficiently characterize the MIMO systems in complex space and get better propagation performance than the conventional MIMO systems considering only free space, which is important for guiding the power and phase allocation of antenna units. In this manuscript, an Electromagnetic-Information-Theory (EMIT) based model is developed for efficient characterization of MIMO systems in complex space. The group-T-matrix-based multiple scattering fast algorithm, the mode-decomposition-based characterization method, and their joint theoretical framework in complex space are discussed. Firstly, key informatics parameters in free electromagnetic space based on a dyadic Green's function are derived. Next, a novel group-T-matrix-based multiple scattering fast algorithm is developed to describe a representative inhomogeneous electromagnetic space. All the analytical results are validated by simulations. In addition, the complete form of the EMIT-based model is proposed to derive the informatics parameters frequently used in electromagnetic propagation, through integrating the mode analysis method with the dyadic Green's function matrix. Finally, as a proof-or-concept, microwave anechoic chamber measurements of a cylindrical array is performed, demonstrating the effectiveness of the EMIT-based model. Meanwhile, a case of image transmission with limited power is presented to illustrate how to use this EMIT-based model to guide the power and phase allocation of antenna units for real MIMO applications.

## I. INTRODUCTION

TYPICALLY, for antenna design, it is promising to maximize the channel capacity via a multiple-input-multiple-output (MIMO) system to approach the limit of channel capacity during propagation. Thus, Under the demand for high accuracy and low latency nowadays, the basic research on efficient characterization of MIMO systems is very important [1]. On this basis, we can carry out further work such as the optimization solutions for the power and phase allocation of antenna units.

Previous works for MIMO characterization can be roughly clarified into two categories: electromagnetic (EM) methods and information theory. The former mainly focuses on the radio-frequency (RF) front-end design by solving Maxwell's equations under different boundary conditions, consisting of the descriptions of the complex electromagnetic space [2], [3], while the latter mainly analyzes the channel properties under different probability models by using Shannon information theory [4], [5]. The above two frameworks are faced with a major challenge in practical application: how to efficiently model MIMO systems in complex space to achieve better propagation performance than MIMO analysis that only consider free space.

For EM methods, the core step of intelligent designs nowadays is reconstructing the MIMO systems' radiation patterns [6], [7], [8]. To consider the effect of the EM propagation space, full-wave numerical algorithms have been used to incorporate the RF front-end design and environment perception into the EM framework, consuming a lot of time [9]. To greatly reduce the calculation time of modeling EM space, some studies have proposed to use approximate methods like ray tracing (RT) [10], [11], [12]. However, this is often not acceptable due to lack of high accuracy. Besides, the T-matrix can be easily used to characterize efficient MIMO in complex EM space, via combining multiple scattering equations [13], [14]. However, practical wireless communication often focuses on some informatic parameters (such as channel capacity), while the EM-only framework is incapable of efficiently extracting the informatic parameters in the complex EM space.

For information theory, the common statistic model, such as the Rayleigh fading model, is a mathematical tool based on the assumption of rich scattering [15]. When it evolves to cluster models like geometry-based stochastic models (GBSMs), the EM space is equivalent to the clusters with different shapes or distributions for convenient characterization [16], [17]. However, the accuracy of those models will be reduced due to the EM properties of the MIMO system. For example, the work in [18] complements numerical methods to make up for the problem of using only Fresnel approximation in airborne antenna design. Moreover, the main idea of the emerging intelligent reflective surface (IRS) is to lay out the controllable

This project is supported in part by Natural Science Foundation of China (NSFC), Grant No. 62071424, 62201499 and 62027805. (Corresponding Author: Da Li, li-da@zju.edu.cn)

The authors are with ZJU-UIUC Institute, Zhejiang Provincial Key Laboratory of Advanced Microelectronic Intelligent Systems and Applications, and the College of Information Science and Electronic Engineering, Zhejiang University, Hangzhou 310027, China.

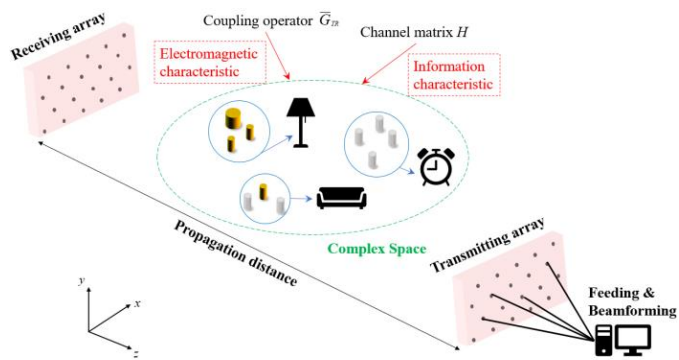

Fig. 1. System model of MIMO analysis in a complex space for indoor communication.

surfaces in free or complex EM space [19], [20], [21], [22], [23]. Due to the lack of efficient MIMO characterization, this technology is still in the trial stage. This suggests that many basic assumptions of the information-only framework need to be reconsidered.

Nowadays, the electromagnetic information theory (EMIT) for the MIMO characterization attracts more attentions, which is expected to solve the challenges mentioned above [24], [25], [26], [27]. Researchers point out that with the wide layout of the antenna array (e.g., Internet of vehicles), it is expected to eliminate the step of channel estimation with the help of rich environmental information [28]. Some works have been done from the perspective of EM fields to study the degree of freedom of MIMO systems [29], [30]. There are also mathematical methods to model the source region and field region as two sets of orthogonal bases in Hilbert space, and then construct some characteristic parameters of the MIMO system [31], [32]. To integrate the RF front-end design in the EMIT framework, the surface currents of antenna elements are modeled as the point sources with orthogonal bases. For example, a model was established to build a channel matrix from the angle of coordinate transformation and orthogonal decomposition of EM plane wave expansion, applied in holographic MIMO system [33], [34]. Additionally, the work in [35] contains the idea of deriving the channel limit of a MIMO system by the EM field method. Nevertheless, the above research works on EMIT mainly focus on free space or revealing the parameter mapping between two theories; efficient characterization algorithms and clear EM information analysis methods for complex EM space are still unexplored.

In this paper, we develop an EMIT-based model to conduct the efficient characterization for MIMO systems in complex EM space. The proposed EMIT-based model uses the group T matrix algorithm and dyadic Green's function-based mode analysis method, filling the research gap of efficient characterization algorithms and clear EM information analyses. The main contributions of this paper are described as follows.

1) The key parameters of the MIMO systems are extracted through the dyadic Green's function and matrix mode analysis. The information characteristics of the MIMO systems are described by the EM method, revealing some important conclusions and deducing the key informatic parameters and valuable conclusions of information theory by means of EM methods.

2) A fast algorithm based on the group T matrix is developed to model the complex EM space. Since the algorithm has semi-analytical characteristics and the classical T matrix can be stored, which provides a faster calculation compared with the traditional full-wave algorithm. In contrast to the RT and pilot-based methods for channel estimation, our EM algorithm can be easily integrated into EMIT due to its higher accuracy and efficiency. In other words, the benefit of our proposed method is generated from the fast characteristics of the group T matrix and the EM analysis of the channel matrix (without the help of statistics).

3) The efficient EMIT-based model is proposed to characterize the MIMO systems in complex space. As a proof-of-concept, a microwave anechoic chamber measurement of a cylindrical array is taken as an example, demonstrating the effectiveness of the EMIT-based model for the MIMO mode analysis. Meanwhile, a case of image transmission with limited power is presented to illustrate how to guide the MIMO feeding based on the model, bringing a new insight into extracting information parameters using the basis of computational electromagnetism.

This article is organized as follows. The key informatics parameters based on the dyadic Green's function are derived in Section II. Then, the proposed EMIT-based model is analyzed in Section III. Experimental verification and an image transmission case were conducted in Section IV. Finally, the conclusion is drawn in Section V.

## II. SYSTEM MODEL AND KEY PARAMETERS

As shown in Fig. 1, consider a typical MIMO system including the transmitting and receiving array for indoor communication, whose overall communication performance will be affected by the propagation distance and the properties of complex space. In this section, the EM propagation space is designated as a free space for extracting key parameters of a MIMO system. More complex EM space is characterized in the next section.

To combine the coupling operator $\overline{\mathbf{G}}_{TR}$ and the channel matrix $\mathbf{H}$, a series of isotropic point sources are placed in the transmission volume and the receiving volume, with the position vectors $\mathbf{r}_T$ and $\mathbf{r}_R$ respectively. The EM wave received is defined as $\mathbf{\psi}_{outR}$, thus the Helmholtz wave equation is given by

$$\nabla \times \nabla \times \mathbf{\psi}_{outR} - k_0^2 \mathbf{\psi}_{outR} = i\omega\mu_0 \mathbf{J}_{incT}, \quad (1)$$

where $\mathbf{J}_{incT}$ is the transmitted source and $k$ is the wave vector. To solve this equation, the dyadic Green's function $\overline{\mathbf{G}}$ operator based on the impulse function idea is introduced:

$$\overline{\mathbf{G}}(\mathbf{r}_R, \mathbf{r}_T) = \left( \overline{\mathbf{I}} + \frac{\nabla\nabla}{k^2} \right) \frac{\exp[ik|\mathbf{r}_R - \mathbf{r}_T|]}{4\pi|\mathbf{r}_R - \mathbf{r}_T|}, \quad (2)$$

where $\bar{\mathbf{I}}$ is the unit tensor. Since the dyadic Green's function tensor $\bar{\mathbf{G}}$ contains the scalar Green's functions:

$$\bar{\mathbf{G}} = \begin{bmatrix} Gxx & Gxy & Gxz \\ Gyx & Gyy & Gyz \\ Gzx & Gzy & Gzz \end{bmatrix}. \quad (3)$$

When it comes to the two independent single-polarization situation at far field, the coupling operator is able to be simplified into the scalar Green's function without loss of accuracy. Therefore, the element $h_{ij}$ in the channel matrix $\mathbf{H}$ will changed to the following form in the case of a single polarized source:

$$h_{ij} = \frac{\exp[ik|\mathbf{r}_{Ri} - \mathbf{r}_{Tj}|]}{4\pi|\mathbf{r}_{Ri} - \mathbf{r}_{Tj}|} = g_{0(ij)}, \quad (4)$$

where $g_0$ is the scalar Green's function, that is, the special form of $\bar{\mathbf{G}}$ in the case of single polarization. It is worth mentioning that the channel matrix $\mathbf{H}$ at this time is not normalized, so it contains the path loss.

Assume that the number of transmitting source points is $N_T$, and the number of receiving field points is $N_R$. Therefore, the transmitting source can be expressed as a $N_T*1$ matrix, the receiving electric field as a $N_R*1$ matrix, and the coupling operator $\bar{\mathbf{G}}$ of the EM space as a $N_R*N_T$ matrix. By introducing the Dirac notation, the MIMO propagation relation of free space is expressed as:

$$|\psi_{outR}\rangle = \bar{\mathbf{G}}|\mathbf{J}_{incT}\rangle. \quad (5)$$

To normalize the channel matrix, we defined the normalized coupling operator $\bar{\mathbf{G}}_{TR}$ as $\bar{\mathbf{G}}_{TR} = \alpha\bar{\mathbf{G}}$, where $\alpha$ is a normalization factor, making $\mathrm{E}\left[\|\bar{\mathbf{G}}_{TR}\|_F^2\right] = N_T N_R$. $\mathrm{E}[\cdot]$ denotes the expectation and $\|\cdot\|_F$ means the Frobenius norm. The physical meaning of this normalization is that every sub-channel should have a unity average channel gain.

According to the Hermitian nature of $\bar{\mathbf{G}}_{TR}\bar{\mathbf{G}}_{TR}^\dagger$, singular value decomposition (SVD) of the coupling operator could conduct mode analysis of MIMO EM propagation, where † denotes the conjugate transpose:

$$\bar{\mathbf{G}}_{TR} = \mathbf{U}_R \mathbf{S} \mathbf{V}_T^\dagger, \quad (6)$$

where $\mathbf{V}_T$ ($\mathbf{U}_R$) is a $N_T*N_T$ ($N_R*N_R$) matrix, and each column represents the EM eigenvector of the transmitting sources (receiving fields). Because of the unitary nature of the SVD eigenmatrix, it is known that each column is strictly orthogonal, which is called the EM space mode. The weight of each pattern is determined by the corresponding element in the diagonal matrix $\mathbf{S}$. The combinations of those orthogonal modes form two Hilbert spaces, and therefore the coupling operator $\bar{\mathbf{G}}_{TR}$ builds a mapping between the transmitting Hilbert space and the receiving Hilbert space, which is an important property in the subsequent discussion.

As we all know, the upper limit of information transmission per bandwidth in MIMO systems is also limited by Shannon's formula [36]:

$$\begin{aligned} C &= \mathrm{E}\left\{\log_2\left[\det\left(\mathbf{I} + \frac{\rho}{n}\frac{1}{N_t}\bar{\mathbf{G}}_{TR}\bar{\mathbf{G}}_{TR}^\dagger\right)\right]\right\}, \\ &= \sum_i \log_2\left(1 + \frac{\rho}{n}\sigma_i^2\right) \end{aligned} \quad (7)$$

where $\mathbf{I}$ is the identity matrix and $\sigma_i$ are the singular values of $(1/\sqrt{N_T})\bar{\mathbf{G}}_{TR}$. Apparently, $\sigma_i^2$ is the decisive parameter of key information-carrying capacity in the MIMO system at a given SNR $\rho/n$. Besides, we can drop the expectation $\mathrm{E}[\cdot]$ in (7) and no longer need to make a special distinction for large-scale and small-scale path loss and fading, because the amplitude and phase changes of the electric field have been included in the operator $\bar{\mathbf{G}}_{TR}$.

It is seen from (6) and (7) that the singular value of EM propagation space determines the number and weight of independent modes, which establishes a corresponding relationship with the number of independently available channels and path loss of wireless communication. We give the key informatics parameters of a MIMO system by referring to the effective rank idea of existing work [21]:

$$C_{eff} = \prod_{i=1}^{\min(N^R,N^T)} \exp(-\sigma_i' \ln(\sigma_i')), \quad (8)$$

where $C_{eff}$ represents the EM effective capacity, $\sigma_i' = \sigma_i/(\sum \sigma_i)$ represents the normalized singular values of $\bar{\mathbf{G}}_{TR}$. Hence, (8) establishes the mapping relationship between the dyadic Green's function matrix and typical informatics parameters, which is an important tool for the MIMO mode analysis.

To understand how this approach works, both mathematically and physically, we set up an $N*N$ MIMO system with the same EM space properties, as shown in Fig. 1. In fact, in practical engineering applications, the mutual coupling is concerned not because it affects the EM equivalent capacity, but because it affects the radiation efficiency and signal-to-noise ratio of the antennas. The transmitting and receiving antennas are modeled as isotropic point sources/receivers (delta function basis), which is a widely used assumption in EM information theory. From the EM perspective, the antennas can also be modeled as continuous surface (equivalent) currents by the rooftop or Rao–Wilton–Glisson (RWG) basis, as frequently utilized in the methods of moments. Different basis representations of the currents, related to different antenna designs, will not influence the estimations of the effective degree of freedom limit. Since we want to focus the analysis in this work on solving dyadic Green's function and extracting informatics parameters in a complex space, we choose the model carefully to avoid mutual coupling. To construct the basic framework of EMIT, three key parameters (the number of sources, communication distance, and the size of antenna aperture) are considered to illustrate the relationship between RF front-end devices' design and the effective capability of the MIMO system.

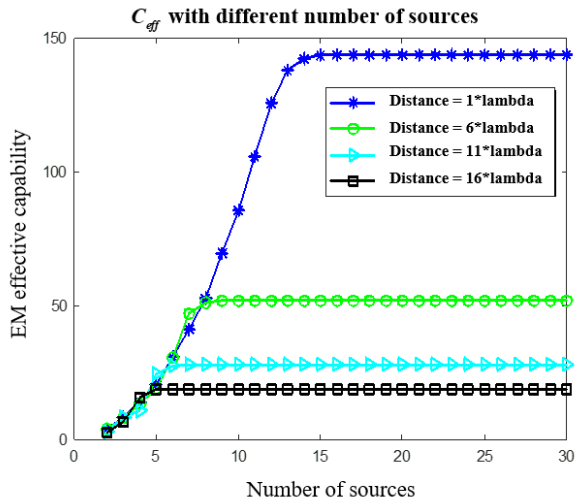

Fig. 2. EM effective capability with the change of number of sources in four communication distances.

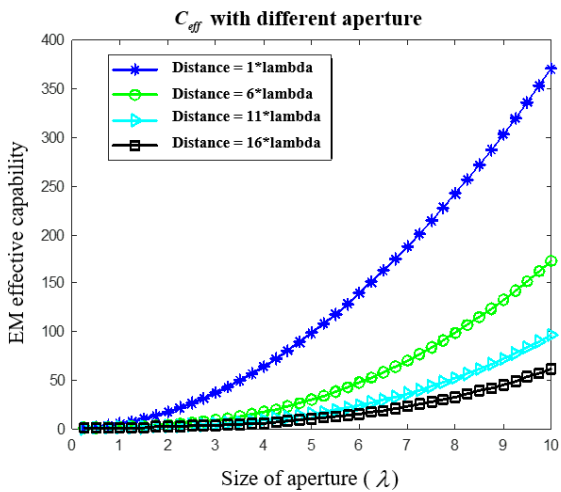

Fig. 3. EM effective capability with the change of aperture sizes in four communication distances.

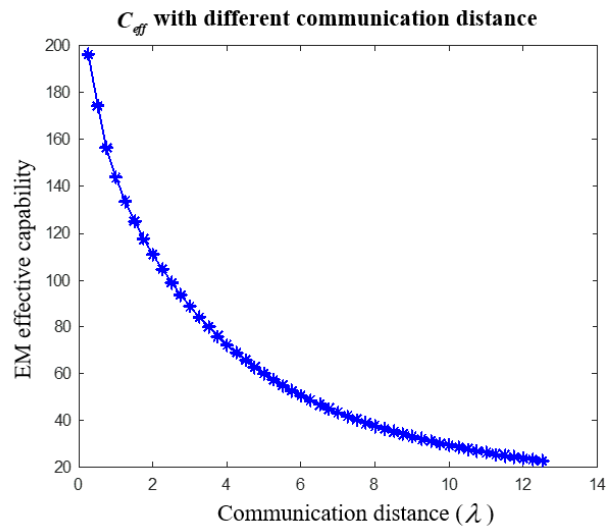

Fig. 4. The attenuation of EM effective capability with the change of communication distance.

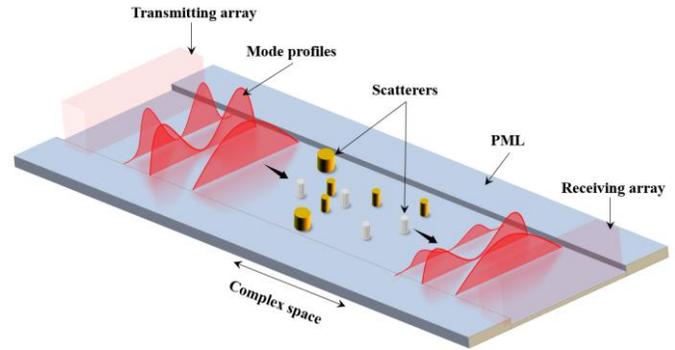

Fig. 5. Schematic diagram of MIMO mode analysis of complex space. The material, position, quantity and shape of the scatterers can be set arbitrarily in our algorithm.

In Fig. 2, the relationship between the EM effective capacity and the number of sources is presented at a given aperture ($6\lambda * 6\lambda$), showing clearly that with the increase of $N$, the EM effective capacity under different communication distances will increase with the same slope, but it converges to the channel capacity. In this case, considering that the change of the total power of the transmitting array will lead to different channel capacities, we fixed the total transmitting power at $P_0$, satisfying $\sum_{i=1}^{N_T} P_i = P_0$.

In addition, to illustrate the physical nature of the convergence, Fig. 3 shows the EM effective capacity corresponding to different aperture sizes with enough point sources (30*30). Obviously, the size of the aperture plays a determinant role in the information capacity of MIMO systems. It suggests that the trend of antenna miniaturization is the weakening of maximum carrying information, which cannot be solved by multi-antenna technology. Besides, in Fig. 4, we plot the curve of EM effective capacity changing with the communication distance, revealing the characteristics of wireless communication-energy attenuated with the propagation distance from the perspective of dyadic Green's function. Besides, in Fig. 3 and Fig. 4, the variables we focus on are the aperture and distance respectively, so the number of point sources is a constant, and the total power $P_0$ always remains a constant.

It is worth mentioning that, due to the basis function decomposition method (such as Rao-Wilton-Glisson (RWG) basis in MoM) commonly used in computational electromagnetics, the specific RF front-end structure can be decomposed into the sum of point sources through grid partitioning, and the multi-channel coupling effect will be considered in the coefficient term of the operator $\overline{\mathbf{G}}_{TR}$. Therefore, when using the above method to perform theoretical modeling of EMIT, the coupling can be characterized by adding a coefficient term to the operator $\overline{\mathbf{G}}_{TR}$, and the specific physical dimensions of the RF front-end can be numerically quantified by base function equivalence.

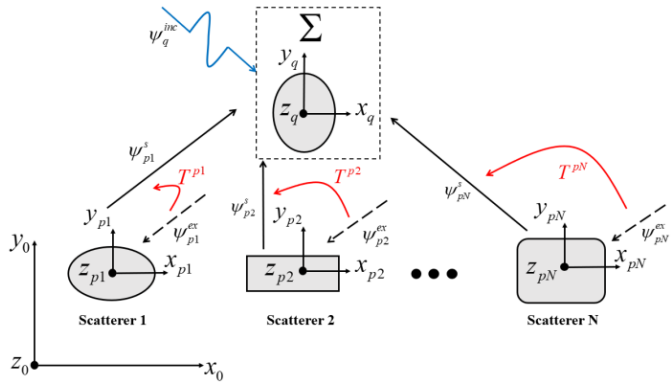

Fig. 6. Illustration of the group-T-matrix-based algorithm. The distribution of the field around the scatterer is decomposed, and the steady-state coefficient matching is carried out based on the cylindrical wave expansion without meshwork and time-domain iteration.

Essentially, changing the RF front-end or the channel will affect the value of $C_{eff}$ in (8) by affecting the distribution of $\sigma_i$ on the $i$th channel. In other words, $C_{eff}$ and the distribution of $\sigma_i$ are the inherent property of the communication system.

However, the core assumption of this part is based on free EM space, and the specific form of coupling operator $\overline{\mathbf{G}}_{TR}$ will change when numerous scatterers are introduced. The next section will demonstrate the fast algorithms for characterizing the complex EM complex space.

## III. Proposed EMIT-Based Model for EM Complex Space

In some typical wireless communication scenarios, objects in complex scattering environments are usually represented by some types of scatterers for convenient EM calculations, among which one of the commonly-used classical models is the cylindrical array, as described in Fig.5. For example, a vehicle-to-vehicle channel is equivalent to a scattering cluster in the internet of vehicles channel modeling [16]. Due to the poor accuracy and long response time of traditional channel measurement schemes, this section proposes an EMIT-based model for efficient MIMO characterization in this typical scattering complex space based on the group T matrix.

### A. Algorithm Description

$N$ cylindrical scatterers in MIMO EM propagation space are considered, which are centered at $r_p (p=1,2,...,N)$ and are with radius $a_p (p=1,2,...,N)$. These parameters can be easily substituted to simulate different distributions and different shapes of scatterers. For the description of the RF front-end, we use a $N_s*1$ dipole antenna array, coordinate $r_s (s=1,2,...,N)$, as a convenient MIMO model. The overall algorithm framework is shown in Fig. 6. To be clear, we focus on the scenarios where the transceivers and receivers are in the same horizontal plane (such as vehicle-to-vehicle communication and indoor point-to-point communication). In this case, we can regard the scatterer as a cluster of cylindrical scatterers, so conducting cylindrical wave expansion is reasonable and convenient. This benefits the convenience of calculation and the simplicity of the model.

To take the coupling between scatterers into account, we take the $q$th scatterer as the analysis object and decompose the total external field $\psi_q^{ex}$ around it into the sum of the incident field $\psi_q^{inc}$ and the scattering field $\psi_p^s$ of the rest scatterers:

$$\psi_q^{ex} = \psi_q^{inc} + \sum_{\substack{p=1 \\ p \neq q}}^{N} \psi_p^s. \tag{9}$$

For solving the scattered fields, the electric field is expanded as a vector cylindrical wave harmonic function:

$$\psi_q^{ex} = \sum_n I_n^{(q)} Rg\psi_n(k(\mathbf{r}-\mathbf{r_q})), \tag{10}$$

where $k$ is the wave vector, $I_n^{(q)}$ is the cylindrical wave coefficient, which is the unknown core quantity for solving the field distribution.

In addition, the specific mathematical form of cylindrical wave expansion in (10) is given as follows:

$$\begin{aligned}\psi_n(k(\mathbf{r}-\mathbf{r_p})) &= H_n^{(1)}(k|\mathbf{r}-\mathbf{r_p}|)\exp(in\phi_{\mathbf{rr_p}}) \\ Rg\psi_n(k(\mathbf{r}-\mathbf{r_p})) &= J_n(k|\mathbf{r}-\mathbf{r_p}|)\exp(in\phi_{\mathbf{rr_p}})\end{aligned}, \tag{11}$$

where $\mathbf{r}$ represents the coordinate vector of the field point, $\phi_{\mathbf{rr_p}}$ represents the angle between the vectors $\mathbf{r}$ and $\mathbf{r_p}$, $J_n$ is the Bessel function of order $n$, $H_n^{(1)}$ is the Hankel function of order $n$, and $Rg$ means regularization. Later, we will use the symbol $i$ to represent the imaginary unit.

For the mode matching, we perform the same cylindrical wave expansion for the incident field $\psi_q^{inc}$ determined by the MIMO RF front-end (here is the $N_s*1$ dipole array), obtaining:

$$\psi_q^{inc} = \sum_{s=1}^{N_s} \frac{i}{4} H_0^{(1)}(k|\mathbf{r}-\mathbf{r_s}|). \tag{12}$$

To obtain the same expansion form as (10), the vector addition theorem is used to further expand (12) to obtain:

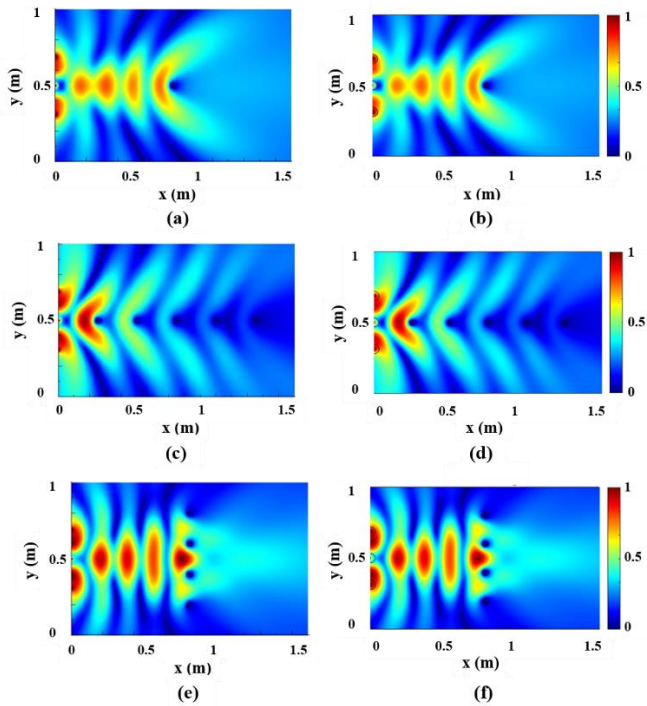

Fig. 7. Normalized electric field distribution on the validation plane. (a) (c) (e): FDTD solver for scatterers distribution of 1*1, 1*5 and 4*1 respectively. (b) (d) (f): proposed EMIT-based model for scatterers distribution of 1*1, 1*5 and 4*1 respectively.

$$\psi_q^{inc} = \frac{i}{4} \sum_n \sum_{s=1}^{N_s} H_n^{(1)}\left(k|\mathbf{r}-\mathbf{r_s}|\right)\exp\left(-in\phi_{\overline{r_s r_q}}\right) \times Rg\psi_n(k(\mathbf{r}-\mathbf{r_s})). \quad (13)$$

In (13), as the RF front-end information is part of prior knowledge, the expansion coefficient of $\psi_q^{inc}$ is determined, which is convenient for solving $I_n^{(q)}$ in (10). Next, we write the scattering field $\psi_p^s$ of the $p$th scatterer as follows:

$$\psi_p^s = \int_{dS_n} \left[\overline{\mathbf{G}}_{TR} \cdot i\omega\mu_0 \mathbf{J}(\mathbf{r_p}') - \nabla \times \overline{\mathbf{G}}_{TR} \cdot \mathbf{M}(\mathbf{r_p}')\right]dS', \quad (14)$$

where $\overline{\mathbf{G}}_{TR}$ is dyadic Green's function illustrated in (2), $\omega$ is the angular frequency, $\mathbf{J}(\mathbf{r_p}')$ and $\mathbf{M}(\mathbf{r_p}')$ are the current density and magnetic current density at the $p$th scatterer, respectively. The EM variation in the complex space is described by the action of the coupling operator $\overline{\mathbf{G}}_{TR}$ on $\mathbf{J}(\mathbf{r_p}')$ and $\mathbf{M}(\mathbf{r_p}')$. When $\mathbf{J}(\mathbf{r_p}')$ and $\mathbf{M}(\mathbf{r_p}')$ do not exist, (9) will then degenerate into the free space case shown in section II.

In order to solve the $\psi_p^s$ described in (14), we use the consistent mathematical form of $\psi_q^{ex}$ on different scatterers to expand $\psi_p^s$ into cylindrical waveform by using (10), and the transformation relationship is shown in the red curve in Fig. 6.

Thus, (14) can be rewritten as follows based on the group-T-matrix:

$$\psi_p^s = \sum_m T_m^{(p)} I_m^{(p)} Rg\psi_m(k(\mathbf{r}-\mathbf{r_s})), \quad (15)$$

where $T^{(p)}$ is the group-T-matrix representing the relationship between the incident field and scattering field of the $p$th clustered scatterer, and its characteristics are only related to the shape and material of the current scatterer. Assuming the internal wave vector of the scatterer is $k_p$, the general form of group-T-matrix in the cylindrical coordinate system can be obtained by using analytical methods:

$$T_m^{(p)} = \frac{k_p J_m'(k_p a_p) - J_m(k_p a_p)kJ_m'(ka_p)}{kH_m^{(1)}(ka_p)J_m(k_p a_p) - H_m^{(1)}(ka_p)k_p J_m'(k_p a_p)}. \quad (16)$$

The T-matrix of any shape objects can be solved by numerical methods such as the method of moments (MoM) according to (14).

The basic purpose of this section is to illustrate the algorithm's efficiency, and thus we consider the model of dipole array with TM polarized waves incident on a perfect electric conductor (PEC). In this case, (16) evolves into:

$$T_m^{(p)} = -\frac{J_m(ka_p)}{H_m^{(1)}(ka_p)}. \quad (17)$$

Substitute (17) into (15) to obtain the field distribution with $I_m^{(p)}$ as the only variable. The matrix equation of the unknown coefficient $I_m^{(p)}$ can be obtained by combining (9), (10), (13), and (15):

$$\mathbf{Z} \cdot \mathbf{I} = \mathbf{V}. \quad (18)$$

Here, in order to solve the coefficient $I_m^{(p)}$, the equations with different scatterers are written in matrix form, and the order of the Bessel function is truncated with the truncation number $N_{\max}$. Therefore, $\mathbf{Z}$ is a square matrix of dimension $(2N_{\max}+1)N$, while $\mathbf{V}$ is a $(2N_{\max}+1)N \times 1$ vector. The specific form is:

$$[\mathbf{Z}]_{(q-1)\times N+n, (p-1)\times N+m} = \begin{cases} -1, & p=q \\ T_m^{(p)} H_{n-m}^{(1)}\left(k|\overrightarrow{r_p}-\overrightarrow{r_q}|\right)\exp\left(-i(n-m)\phi_{\overline{r_p r_q}}\right), & p \neq q \end{cases}, (19)$$

$$[\mathbf{V}]_{(q-1)\times N+n} = -\frac{i}{4}\sum_{s=1}^{N_s} H_n^{(1)}\left(k|\overrightarrow{r_s}-\overrightarrow{r_q}|\right)\exp\left(-in\phi_{\overline{r_s r_q}}\right). \quad (20)$$

**TABLE I**

COMPARISON OF CPU TIME AND RMS ERROR BETWEEN FDTD AND PROPOSED EMIT-BASED MODEL FOR THE CHARACTERISTIC OF COMPLEX SPACE

| Scatterer Distribution (Total Number) | FDTD CPU Time (s) | EIT-Based CPU Time (s) | RMS Error |
|---|---|---|---|
| 1*1 (1) | 2.38 | 0.0104 | 0.86% |
| 1*5 (5) | 2.70 | 0.0449 | 0.52% |
| 4*1 (4) | 3.19 | 0.0268 | 0.18% |
| 4*5 (20) | 5.62 | 0.514 | 1.28% |
| 10*15 (150) | 8.74 | 6.53 | 0.07% |
| Radom (90) | 27.6 | 4.72 | 1.02% |

In (19) and (20), $\mathbf{Z}$ is determined by the properties of the complex space, and $\mathbf{V}$ is determined by the properties of the RF front-end. A joint solution can semi-analytically describe the evolution of the MIMO coupling operator $\overline{\mathbf{G}}_{EIT}$. Therefore, due to the change of EM space coupling operator, (5) will be rewritten as:

$$\left| \psi_{outR} \right\rangle = \overline{\mathbf{G}}_{EIT} \left| \mathbf{J}_{incT} \right\rangle. \quad (21)$$

The subsequent analysis only needs to be carried out in the same way as (6)-(8) to complete efficient MIMO characterization in complex space. It should be noted that, when facing the time-varying channel scenario, it will be very convenient to rewrite the coupling operator $\overline{\mathbf{G}}_{EIT}$ into the form based on the time-domain Green's function.

### B. Numerical Results

To verify the accuracy and efficiency of the proposed EMIT-based model, numerical calculations of some specific scenarios are carried out and compared with full-wave simulation results.

Fig. 7 presents the field distributions of three simple arrays. The effectiveness of the EMIT-based model is verified by comparing the full-wave FDTD algorithm (a, c, e) with the proposed algorithm (b, d, f). We consider an EM space of 1 m*1.5 m, where the MIMO system is modeled as a 3*1 dipole array with an aperture of 0.75 m. The scatterer array element is modeled as a metal cylinder with a height of 0.25 m and a radius of 0.015 m, and the boundary is set as PEC. Due to the dense mesh division of the full-wave algorithm, its application is severely limited. However, the proposed semi-analytic algorithm based on group-T-matrix is suitable for various frequencies because it does not need mesh division. To obtain the comparison results, we first define the operating frequency at 915 MHz in this section.

Fig. 7 illustrates that the proposed EMIT-based model has achieved good results and can accurately describe the complex space. In order to further demonstrate its efficiency, the scatterer distribution was adjusted, and the solving time and error of the EMIT-based model and FDTD were calculated by analyzing the field intensity curve at the RF back-end, as shown in Table I. It is worth noting that the 10*15 distribution cannot fully explain the difference between the two algorithms, because the large number of FDTD meshes converge extremely fast due to the inability of the electric field to propagate effectively, and the solutions are often mediocre at this time. Therefore, we further consider the case of a random array, that is, randomly removing 60 scatterers from the 10*15 scatterer distribution. Besides, we clarify that the running time of our proposed algorithm mainly depends on the number of scatters. Therefore, the proposed EMIT-based model has higher computational efficiency than full-wave algorithms like FDTD, which provides great convenience for the description of complex space. But generally, the complexity of the real-world environment increases with the communication distance. In this case, an efficient way to leverage the EMIT-based method is to build a common clustering model database. Compared with pilot-based methods, it also has good efficiency under the condition of a complete database. For example, a vehicle-to-vehicle channel is equivalent to a scattering cluster in the internet-of-vehicles channel modeling [16].

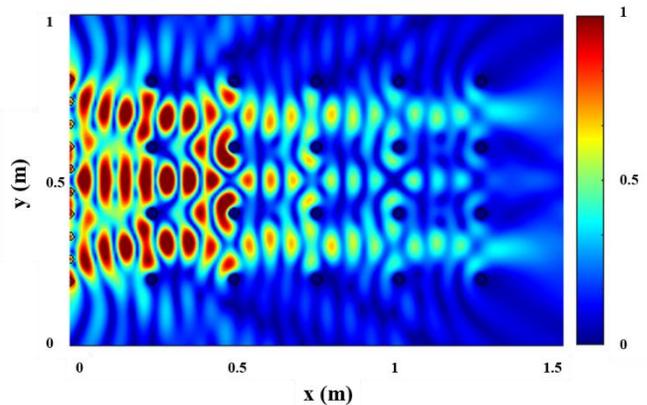

Fig. 8. The total normalized electric field distribution on the validation plane corresponding to the proposed complex space obtained by EMIT-based model.

### C. Mode Analysis Step of the EMIT-Based Model

After efficient characterization of the complex space is verified, the EMIT-based model performs a mode analysis of the above characterization results to obtain theoretical interpretations to guide the design of wireless communications.

Consider an actual information transmission scenario where the RF front-end is a single-polarized dipole antenna array operating at 2.5GHz (equivalent using an ideal line source operating at 2.5Ghz), with a scale of 10*1 (designed to make the MIMO feature more obvious), and the complex space is simplified to a 4*5 metallic cylindrical scatterer cluster. According to the quick algorithm in the previous section, the electric field distribution on the validation plane is shown in Fig. 8. By substituting the solved coupling operator $\overline{\mathbf{G}}_{EIT}$ into (6), the EM effective capacity $C_{eff}$ of this model in wireless communication is known as 5.2, which means that the actual effective number of available channels is 5. However, Fig. 3 shows that dyadic Green's function operators $\overline{\mathbf{G}}_{TR}$ (coupling operators in free space) will bring channel gains far beyond 5.2. Therefore, the EMIT-based model provides a convenient tool to quantitatively explain the influence of the complex environment on wireless communication quality. The information at the receiving end in Fig. 8 can help us get the operator $\overline{\mathbf{G}}_{EIT}$. Through the SVD mentioned above, 10 modes at the transmitting end can be decomposed, and the EM responses of these 10 modes in the complex space are shown in Fig. 9. It is clearly found that the first five modes successfully send signals to the receiver effectively in different coupling paths. However, the coupling paths of higher-order modes bypass the receiver's acceptance range and become unavailable modes in wireless communication.

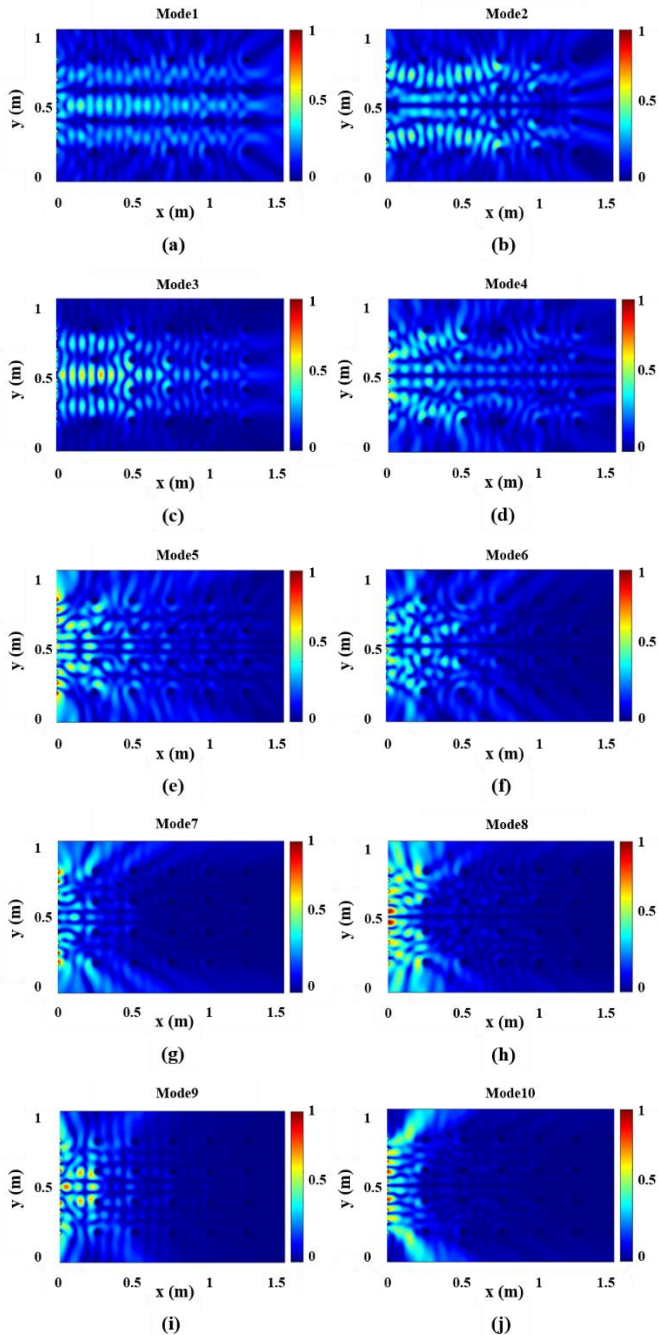

Fig. 9. Each mode's normalized electric field diagram on the validation plane obtained by EMIT-based model. (a-e) Available information transfer modes; (d-j) Unavailable higher-order information transfer modes.

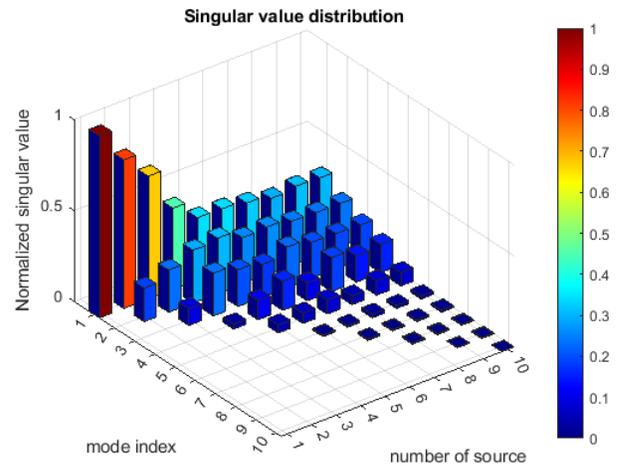

Fig. 10. The distribution of normalized singular values of different number of sources.

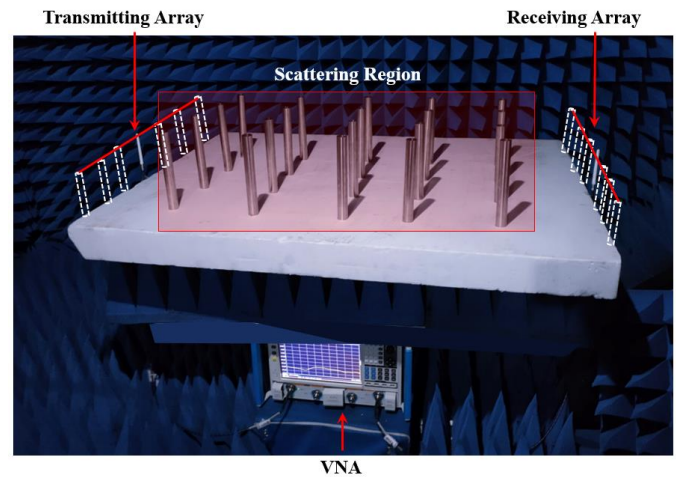

Fig. 11. Simple MIMO propagation system in complex space.

To define the concepts of "available" and "unavailable" more clearly, we show the distribution of modes' singular values for the different number of channels in Fig. 10. A formal definition is given as follows: if all modes are numbered according to the normalized singular value in a descending order like Fig. 10, then the available modes are defined as those whose index is less than the EM effective capacity $C_{eff}$, and the other modes are defined as the unavailable modes. Since the power resources in an actual wireless communication system are limited, the mode weight corresponding to each channel number is normalized here. Obviously, for a MIMO system, there will be an evident truncation of the modes' singular value distribution, and modes below the truncation usually become "unavailable". The number of "available" modes will be strictly determined by (8) after the coupling operator $\overline{\mathbf{G}}_{EIT}$ is obtained by the EMIT-based model.

Obviously, the more channels available, the more information that can be transmitted, and the greater the channel capacity of the corresponding EM space. However, communication resources of the RF front-end are often limited, so it is important to allocate resources properly to achieve better information transmission efficiency. In the next section, power distribution is taken as the background problem to discuss the guidance significance of the EMIT-based model for real wireless communication in a complex environment.

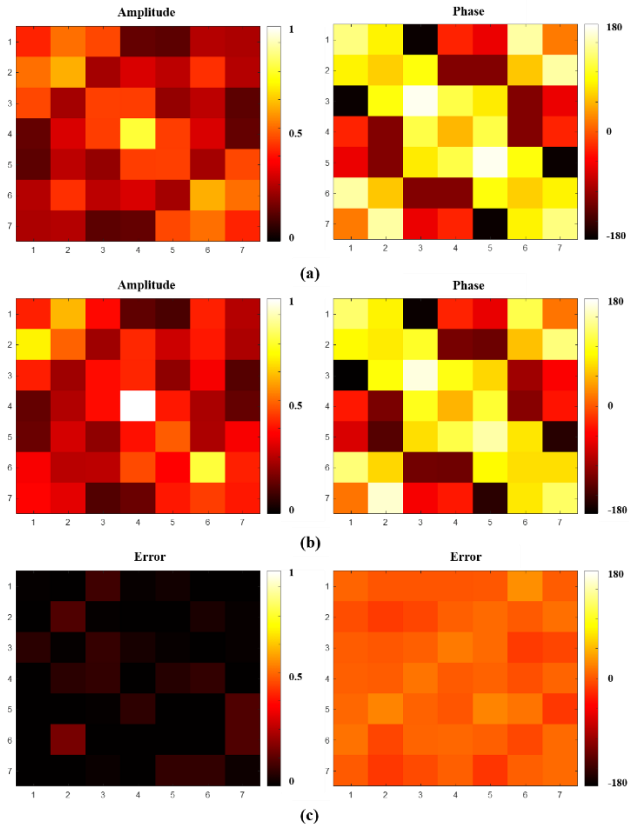

Fig. 12. The normalized amplitude and phase of $S_{21}$ in 7*7 MIMO system. (a) Simulation results in EMIT-based model; (b) Measurement results; (c) The error of the two above.

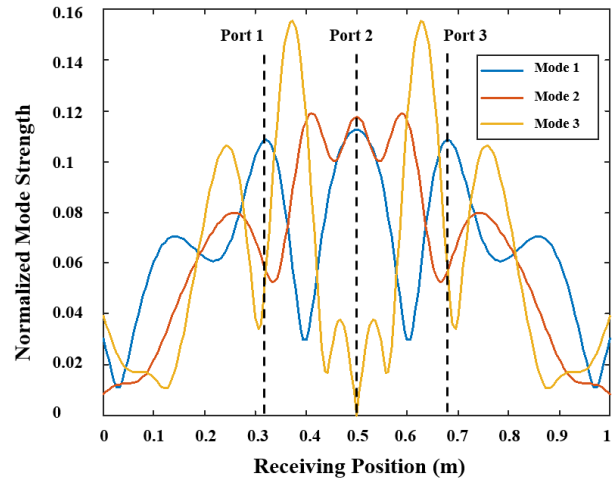

Fig. 13. The field distribution of three orthogonal modes at the receiver aperture. The locations of the three sources are indicated by black dotted lines on the diagram.

## IV. Experimental Analysis

It is worth noting that the above discussion on the application of the EMIT-based model is carried out by simulation. In order to fully explain the effectiveness of the EMIT-based model and the application method under the background of wireless communication, we carried out experimental exploration with the aid of a 3*1 MIMO system.

The experiment construction is shown in Fig. 11, where the system is surrounded by the absorption boundary covered with absorbing materials, and cylindrical metal scatterers with a height of 0.25 m and a radius of 0.015 m are uniformly distributed in the EM space with 4*5 arrays. The transmitting sources and the receiving fields were replaced by dipole antennas with a center frequency of 2.5 GHz and a gain of 2dBi. The vector network analyzer (VNA) is used to measure the $S_{21}$ between transmitting and receiving dipoles through the coaxial feed. Since the measurement of channel matrix elements is concerned with the single excitation properties of MIMO, we replace the actual MIMO system by changing the spatial position of the antenna in the transmitting aperture (shown as the dotted white lines). Therefore, in the experimental design, we selected the weak coupling scenario with the antenna spacing as half-wavelength, and then measured the antenna excitation separately to avoid the impact of coupling on the verification of our EMIT-based model.

For the same scene, we performed effective characterization with the EMIT-based model, and the characterization results are shown in Fig. 12. Fig. 12 (a) and (b) respectively represent the amplitude and phase comparison results between the simulation results of EMIT-based model and the experimental measurement values. The experimental results fully demonstrate the effectiveness of the EMIT-based model in complex space characterization. The purpose of conducting 7*7 channel measurement in our experiment is to verify the EMIT-based model more convincingly. However, the following is mainly to illustrate how the EMIT-based model guides the RF front-end signal transmission. Therefore, to simplify the demonstration process, we select three groups of data evenly spaced to form a new 3*3 MIMO system. In fact, the selection of 3*3 channel positions is arbitrary. However, in this paper, to make the mode orthogonality more significant and avoid the influence of mode crosstalk on the transmitting strategy, three positions with relatively small mode crosstalk are selected, and the crosstalk matrix **CT** is as follows:

$$\mathbf{CT} = \begin{pmatrix} 1 & 0.1857 & 4.03*10^{-15} \\ 0.1857 & 1 & 6.51*10^{-15} \\ 4.03*10^{-15} & 6.51*10^{-15} & 1 \end{pmatrix}. \quad (22)$$

Consider the following problems in an actual wireless communication scenario: 3 * 3 MIMO system needs to conduct data transmission in disorder EM space (with the standard deviation of noise $\chi$), the maximum transmitted power of RF front-end is $P_0$. Under this constraint, since it is a very important subject to consider the optimal power distribution, which is related to whether the upper bound for capacity can be achieved, the coupling operator $\overline{\mathbf{G}}_{EIT}$ obtained by the EMIT-based model is used to endow the shape of the transmitted signal to obtain the best quality of information transmission.

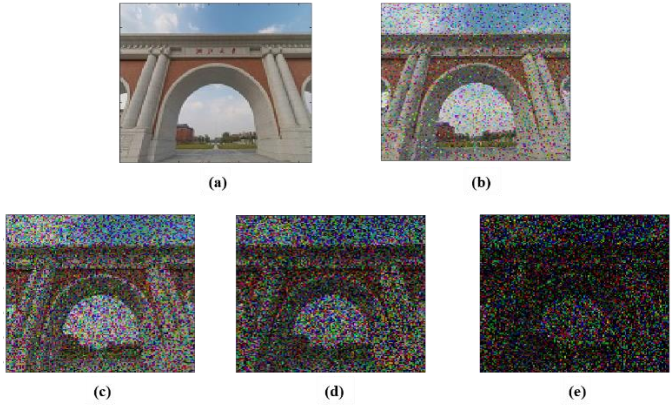

Fig. 14. Image transmission case based on EMIT-based model. (a) Original image with 128*128 pixels. (b) Optimized power distribution. (c)-(e) Conducting single-mode transmission using mode 1, mode 2 and mode 3 respectively.

Rewrite (5-6) to obtain a new coupling equation based on the scenario we're considering:

$$\mathbf{U}_{EIT}^{\dagger}\mathbf{Y} = \mathbf{S}\mathbf{V}_{EIT}^{\dagger}\mathbf{X}, \quad (23)$$

where $\mathbf{U}_{EIT}$ and $\mathbf{V}_{EIT}$ are determined by $\overline{\mathbf{G}}_{EIT}$ to guide the signal processing of transmitter and receiver, respectively. $\mathbf{X}$ and $\mathbf{Y}$ represent the signal form of transmitter and receiver, respectively. The singular value matrix $\mathbf{S}$ is disassembled to obtain the received signal evaluation function $f$:

$$f = \sum_{m=1}^{3}\langle \sigma_m V_m | \alpha \rangle. \quad (24)$$

In (24), $\mathbf{X}$ is decomposed as a bitstream of information X (here we use simple binary phase-shift keying (BPSK) modulation) multiplied by the excitation coefficient $|\alpha\rangle$, and the unitary matrix $\mathbf{V}_{EIT}$ is decomposed into three-mode vectors $|V_m\rangle(m=1,2,3)$. Since there are three sources in our MIMO system, both $|V_m\rangle$ and $|\alpha\rangle$ here have three elements. $\sigma_m(m=1,2,3)$ are the diagonal elements of the singular value matrix $\mathbf{S}$, representing the influence of each mode on the receiving end. To have a clearer understanding of $\sigma_m$, we depicted the electric field distribution at the receiving end with the help of the EMIT-based model, as shown in Fig. 13. The calculated proportions of the three modes are 67.55%, 23.19%, and 9.25%, respectively. Mode 1 with the strongest proportion just contributes its crest to the receiving end, while mode 3 with the weakest proportion just contributes its trough to the receiving end. This provides a clear perspective for signal waveform design from the EM point of view, and reveals that EM space is not the only factor determining mode contribution, and EM space characteristics and RF front-end characteristics should be considered together.

According to the crosstalk matrix calculated in (22), we treat these three modes as orthogonal. Therefore, $|V_m\rangle$ becomes a set of orthogonal basis in a Hilbert space, meeting $\langle V_m^{\dagger}|V_n\rangle = 0$ and $\langle V_m^{\dagger}|V_m\rangle = 1$. Hence, $|\alpha\rangle$ can be written as an orthogonal basis expansion:

$$|\alpha\rangle = \sum_{m=1}^{3}\lambda_m|V_m\rangle, \quad (25)$$

where $\lambda_m$ represent the corresponding weight of each basis vector, which determines the power distribution on the transmitting source. Therefore, the constraint of constant total power $P_0$ can be equivalent to that the excitation vector is located on a fixed circle in the Hilbert space, and the received signal evaluation function $f$ is the sum of the weighted projections of the excitation vector on the three basis functions:

$$\begin{cases} \text{find}: \lambda_m(m=1,2,3) \\ \quad \max: f \\ s.t.: \sum \lambda_m = P_0 \end{cases}. \quad (26)$$

The optimization problem can be easily solved by using Cauchy inequality. By substituting (8), the information transfer function can reach the maximum value only when $\lambda_m/\sigma_m$ is a constant for different $m$. This is similar to the "water-filling" algorithm in channel estimation, while the core difference is that the key informatics parameters in this paper are deduced by an effective EM algorithm. In addition, $\overline{\mathbf{G}}_{TR}\overline{\mathbf{G}}_{TR}^{\dagger}$ or $\overline{\mathbf{G}}_{EIT}\overline{\mathbf{G}}_{EIT}^{\dagger}$ have the same physical meaning as the transmit signal covariance matrix, and the main difference is that $\overline{\mathbf{G}}_{TR}\overline{\mathbf{G}}_{TR}^{\dagger}$ or $\overline{\mathbf{G}}_{EIT}\overline{\mathbf{G}}_{EIT}^{\dagger}$ is calculated by the EM methods based on dyadic Green's function.

It is seen from the mode analysis based on the EMIT-based model that only under the guidance of a specific power allocation strategy, MIMO information transmission can achieve the effect of receiving power equal to transmitting power times path loss. To fully illustrate the guiding significance of the EMIT-based model for power distribution (essentially waveform design), Fig. 14 shows an image transmission case. BPSK is used to discretize every pixel in the picture into an 8-bit data stream for transmission, and noise $\chi$ is joined to EM space. Obviously, under the premise of not processing channel noise, the RF front-end working strategy based on the EMIT-based model is much better than other transmission modes. Therefore, the EMIT-based model can not only efficiently represent the complex space, but also make more valuable guidance for wireless communication.

V. CONCLUSION

In this article, an EMIT-based model is presented to simulate the performance of MIMO systems in complex EM complex space effectively. Firstly, the EM expression of the information coupling operator is given in the free space, and two key informatics parameters, EM effective capacity and path loss, are extracted from the EM perspective. It is proved that the MIMO antennas' aperture is the critical factor in the EM effective capacity of the MIMO system. Secondly, the basic principle of the EM representation method in complex space is given, and several typical scenarios are analyzed, which proves the accuracy and efficiency of the EMIT-based model proposed. The results show that the EMIT-based model can reliably analyze the electromagnetic space about 10% of the time compared to the full-wave simulation. Finally, the MIMO performance in real propagation scenarios is calculated using the EMIT-based model. The experimental results verify that the channel matrix calculated is in good agreement with the measured ones. Based on this, it is pointed out how the EMIT-based model can effectively guide the MIMO design and feeding in a power distribution question.

The ultimate goal of the proposed EMIT-based model is to advance the development of EMIT and demonstrate a new idea of extracting information parameters to the antenna & propagation community using the basis of computational electromagnetism. Currently, it is suitable for cluster models with arbitrary distribution, size, and material, providing an efficient and reliable method for guiding the power and phase allocation of antenna units in scattering complex space. The proposed model can also be easily extended to the guidance of MIMO antenna design in complex spaces by numerical discrete and optimization methods.

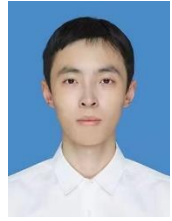

**Jinyan Ma** received the B.S. degree in engineering from Zhejiang University, Hangzhou, China, in 2021. He is currently working toward the Ph.D. degree in electronics science and technology with the College of Information Science and Electronic Engineering, Zhejiang University, Hangzhou, China.

His current research interests include the electromagnetic information theory and efficient electromagnetic calculation methods.

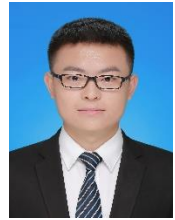

**Zhaoyang Feng** received the B.Sc degree from North China Electric Power University, Beijing, China, in 2017. He is currently working toward the Ph.D. degree in the College of Information Science and Electronic Engineering, Zhejiang University, Hangzhou, Zhejiang.

His current research interests include electromagnetic compatibility, computational electromagnetics and multiple scattering theory

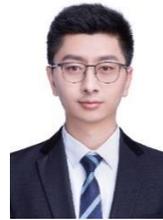

**Ling Zhang** (Member, IEEE) received the B.S. degree in electrical engineering from Huazhong University of Science and Technology, Wuhan, China, in 2015, and the M.S. and Ph.D. degrees from Missouri S&T, Rolla, MO, USA, in 2017 and 2021, respectively, both in electrical engineering. He was with Cisco as a student intern from Aug. 2016 to Aug. 2017. He joined Zhejiang University, Hangzhou, China as a research fellow in 2021. He has authored and co-authored more than 30 journal and conference papers. His research interests include machine learning, power integrity, electromagnetic interference, radio-frequency interference, and signal integrity.

Dr. Zhang was an Organizing Committee, Special Session Chair, Workshop Session Chair, and Poster Session Chair in APEMC 2022. He has given invited presentations at the IBIS Summit at 2021 IEEE Virtual Symposium on EMC+SIPI, and the 2021 Virtual Asian IBIS Summit China. He was the recipient of the Honorable Mention Paper in APEMC 2022, the Best Paper Award in DesignCon 2019, and the Student Paper Finalist Award in ACES Symposium in 2021. He was also the recipient of the Outstanding Young Scientist Reward in APEMC 2022.

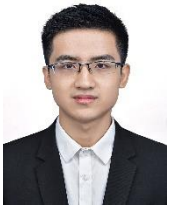

**Ruifeng Li** received the B.S. degree in engineering from University of Electronic Science and Technology of China, Chengdu, China, in 2020. He is currently pursuing the Ph.D. degree at the College of Information Science and Electronic Engineering, Zhejiang University.

His current research interests include the electromagnetic information theory for wireless communication, and efficient calculation methods applied in MIMO antennas.

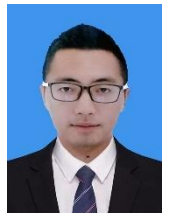

**Da Li** received the B.S. degree in 2014, and the Ph.D. degree in 2019, from Zhejiang University, Hangzhou, China, both in electrical engineering. From 2017 to 2018, he worked at Nanyang Technological University, Singapore, as a Project Researcher. From 2019 to 2021, he joined Science and Technology on Antenna and Microwave Laboratory, Nanjing, China, as a Research Fellow. He is currently an assistant professor at Zhejiang University. His research interests include machine learning, antennas, matesurfaces, and electromagnetic compatibility. Dr. Li has authored or coauthored more than 40 refereed papers and served as Reviewers for 6 technical journals and TPC Members of 3 IEEE conferences. He was also a recipient of the Outstanding Young Scientist Award at 2022 Asia-Pacific International Symposium on Electromagnetic Compatibility.

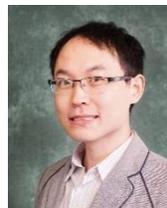

**Shurun Tan** (S'14-M'17) received the B.E. degree in information engineering and M.Sc. degree in electromagnetic field and microwave techniques from the Southeast University, Nanjing, China, in 2009 and 2012, respectively, and the Ph.D. degree in electrical engineering from the University of Michigan, Ann Arbor, MI, USA, in Dec. 2016.

Dr. Tan is an assistant professor in the Zhejiang University / University of Illinois at Urbana-Champaign Institute located at the International Campus of Zhejiang University, Haining, China. He is also affiliated with the State Key Laboratory of Modern Optical Instrumentation, and the College of Information Science and Electronic Engineering, Zhejiang University, Hangzhou, China. He is also an adjunct assistant professor in the Department of Electrical and Computer Engineering, University of Illinois at Urbana-Champaign, Urbana, USA. From Dec. 2010 to Nov. 2011, he was a Visiting Student with the Department of Electrical and Computer Engineering, the University of Houston, Houston, TX, USA. From Sep. 2012 to Dec. 2014, he was a PhD candidate with the Department of Electrical Engineering, the University of Washington, Seattle, WA, USA. From Jan. 2015 to Dec. 2018, he had been affiliated with the Radiation Laboratory, and the Department of Electrical Engineering and Computer Science, the University of Michigan, Ann Arbor, first as a PhD candidate, and then as a postdoctoral research fellow since Jan. 2017.

Dr. Tan is working on electromagnetic theory, computational and applied electromagnetics. His research interests include electromagnetic scattering of random media and periodic structures, microwave remote sensing, electromagnetic information systems with electromagnetic wave-functional devices, electromagnetic integrity in high-speed and high-density electronic integration, electromagnetic environment and reliability of complex electronic systems, etc.

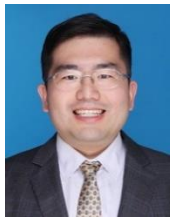

**Wei E. I. Sha** (M'09-SM'17) received the B.S. and Ph.D. degrees in Electronic Engineering at Anhui University, Hefei, China, in 2003 and 2008, respectively. From Jul. 2008 to Jul. 2017, he was a Postdoctoral Research Fellow and then a Research Assistant Professor in the Department of Electrical and Electronic Engineering at the University of Hong Kong, Hong Kong. From Mar. 2018 to Mar. 2019, he worked at University College London as a Marie Skłodowska-Curie Individual Fellow. From Oct. 2017, he joined the College of Information Science & Electronic Engineering at Zhejiang University, Hangzhou, China, where he is currently a tenure-tracked Assistant Professor.

Dr. Sha has authored or coauthored 180 refereed journal papers, 150 conference publications (including 5 keynote talks and 1 short course), 9 book chapters, and 2 books. His Google Scholar citation is 8193 with h-index of 45. He is a senior member of IEEE and a member of OSA. He served as Reviewers for 60 technical journals and Technical Program Committee Members of 10 IEEE conferences. He also served as Associate Editors of IEEE Journal on Multiscale and Multiphysics Computational Techniques, IEEE Open Journal of Antennas and Propagation, and IEEE Access. In 2015, he was awarded Second Prize of Science and Technology from Anhui Province Government, China. In 2007, he was awarded the Thousand Talents Program for Distinguished Young Scholars of China. He was the recipient of ACES Technical Achievement Award 2022 and PIERS Young Scientist Award 2021. Dr. Sha also received 6 Best Student Paper Prizes and one Young Scientist Award with his students.

His research interests include theoretical and computational research in electromagnetics and optics, focusing on the multiphysics and interdisciplinary research. His research involves fundamental and applied aspects in computational and applied electromagnetics, nonlinear and quantum electromagnetics, micro- and nano-optics, optoelectronic device simulation, and multiphysics modeling.

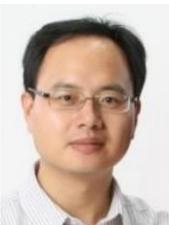

**Hongsheng Chen** received the B.S. and Ph.D. degrees in electrical engineering from Zhejiang University (ZJU), Hangzhou, China, in 2000 and 2005, respectively.

In 2005, he became an Assistant Professor with ZJU, where he was an Associate Professor in 2007 and a Full Professor in 2011.

He was a Visiting Scientist from 2006 to 2008 and a Visiting Professor from 2013 to 2014 with the Research Laboratory of Electronics, Massachusetts Institute of Technology, Cambridge, MA, USA. He is currently a Chang Jiang Scholar Distinguished Professor with the Electromagnetics Academy, ZJU. He has coauthored more than 200 international refereed journal papers. His works have been highlighted by many scientific magazines and public media, including *Nature*, *Scientific American*, *MIT Technology Review*, *Physorg*, and so on. His current research interests include metamaterials, invisibility cloaking, transformation optics, graphene, and theoretical and numerical methods of electromagnetics.

Dr. Chen serves as a Regular Reviewer for many international journals on electromagnetics, physics, optics, and electrical engineering. He serves as a Topical Editor for the Journal of Optics and the Editorial Board for *Nature's Scientific Reports* and *Progress in Electromagnetics Research*. He was a recipient of the National Excellent Doctoral Dissertation Award in China in 2008, the Zhejiang Provincial Outstanding Youth Foundation in 2008, the National Youth Top-Notch Talent Support Program in China in 2012, the New Century Excellent Talents in University of China in 2012, the National Science Foundation for Excellent Young Scholars of China in 2013, and the National Science Foundation for Distinguished Young Scholars of China in 2016. His research work on an invisibility cloak was selected in Science Development Report as one of the representative achievements of Chinese Scientists in 2007.

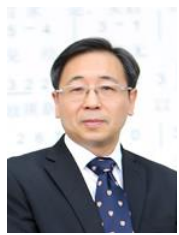

**Er-Ping Li** (S'91, M'92, SM'01, F'08) is currently a Qiushi-Distinguished Professor with Department of Information Science and Electronic Engineering, Zhejiang University, China; served as Founding Dean for Institute of Zhejiang University - University of Illinois at Urbana-Champaign in 2016. From 1993, he has served as a Research Fellow, Associate Professor, Professor and Principal Scientist and Senior Director at the Singapore Research Institute and University. Dr Li authored or co-authored over 400 papers published in the referred international journals, authored two books published by John-Wiley-IEEE Press and Cambridge University Press. He holds and has filed a number of patents at the US patent office. His research interests include electrical modeling and design of micro/nano-scale integrated circuits, 3D electronic package integration.

Dr. Li is a Fellow of IEEE, and a Fellow of USA Electromagnetics Academy, a Fellow of Singapore Academy of Engineering. He is the recipient of IEEE EMC Technical Achievement Award in 2006, Singapore IES Prestigious Engineering Achievement Award and Changjiang Chair Professorship Award in 2007, 2015 IEEE Richard Stoddard Award on EMC, 2021 IEEE EMC Laurence G. Cumming Award and Zhejiang Natural Science 1st Class Award. He served as an Associate Editor for the IEEE MICROWAVE AND WIRELESS COMPONENTS LETTERS from 2006-2008 and for IEEE TRANSACTIOSN on EMC from 2006-2021, Guest Editor for 2006 and 2010 IEEE TRANSACTIOSN on EMC Special Issues, Guest Editor for 2010 IEEE TRANSACTIONS on MTT APMC Special Issue. He is currently an Associate Editor for the IEEE TRANSACTIONS ON SIGNAL and POWER INTEGRITY and Deputy Editor in Chief of Electromagnetics Science. He has been a General Chair and Technical Chair, for many international conferences. He was the President for 2006 International Zurich Symposium on EMC, the Founding General Chair for Asia-Pacific EMC Symposium, General Chair for 2008, 2012, 2016, 2018, 2022 APEMC, and 2010 IEEE Symposium on Electrical Design for Advanced Packaging Systems. He has been invited to give 120 invited talks and plenary speeches at various international conferences and forums.